\documentclass[10pt,journal,compsoc]{IEEEtran}

\usepackage{amsmath,amsfonts}
\usepackage{algorithmic}
\usepackage{algorithm}
\usepackage{array}
\usepackage[caption=false,font=normalsize,labelfont=sf,textfont=sf]{subfig}
\usepackage{textcomp}
\usepackage{stfloats}
\usepackage{url}
\usepackage{verbatim}
\usepackage{graphicx}
\usepackage{multirow}
\usepackage{booktabs}
\usepackage{doi}
\usepackage{threeparttable}
\usepackage[numbers,sort&compress]{natbib}

\ifCLASSINFOpdf
\else
\fi

\begin{document}
%
\title{Investigating Acoustic-Textual Emotional Inconsistency Information for Automatic Depression Detection}
%
%
%
%

\author{Rongfeng Su\IEEEauthorrefmark{1}, Changqing Xu\IEEEauthorrefmark{1}, Xinyi Wu, Feng Xu, Xie Chen,~\IEEEmembership{Member,~IEEE,} Lan Wang\IEEEauthorrefmark{2},~\IEEEmembership{Member,~IEEE,} and Nan Yan\IEEEauthorrefmark{2},~\IEEEmembership{Member,~IEEE}

\IEEEcompsocitemizethanks{\IEEEcompsocthanksitem Rongfeng Su, Lan Wang and Nan Yan are with Guangdong-Hong Kong-Macao Joint Laboratory of Human-Machine Intelligence-Synergy Systems, Shenzhen Institutes of Advanced Technology, Chinese Academy of Sciences, China.
E-mail: \{rf.su, lan.wang, nan.yan\}@siat.ac.cn.
\IEEEcompsocthanksitem Changqing Xu and Xinyi Wu are with the University of Chinese Academy of Sciences, China. E-mail: \{cq.xu, xy.wu2\}@siat.ac.cn.
\IEEEcompsocthanksitem Feng Xu is with the Good Mood Health Industry Group Co., Ltd, China.
E-mail: xufeng@haoxinqing.cn.
\IEEEcompsocthanksitem Xie Chen is with the Shanghai Jiao Tong University, China.
E-mail: chenxie95@sjtu.edu.cn .}

}

\IEEEtitleabstractindextext{%
\begin{abstract}
Previous studies have demonstrated that emotional features from a single acoustic sentiment label can enhance depression diagnosis accuracy.
Additionally, according to the Emotion Context-Insensitivity theory and our pilot study, individuals with depression might convey negative emotional content in an unexpectedly calm manner, showing a high degree of inconsistency in emotional expressions during natural conversations.
So far, few studies have recognized and leveraged the emotional expression inconsistency for depression detection.
In this paper, a multimodal cross-attention method is presented to capture the Acoustic-Textual Emotional Inconsistency (ATEI) information.
This is achieved by analyzing the intricate local and long-term dependencies of emotional expressions across acoustic and textual domains, as well as the mismatch between the emotional content within both domains.
A Transformer-based model is then proposed to integrate this ATEI information with various fusion strategies for detecting depression.
Furthermore, a scaling technique is employed to adjust the ATEI feature degree during the fusion process, thereby enhancing the model's ability to discern patients with depression across varying levels of severity.
To best of our knowledge, this work is the first to incorporate emotional expression inconsistency information into depression detection.
Experimental results on a counseling conversational dataset illustrate the effectiveness of our method.
\end{abstract}

}

\maketitle

{
\renewcommand{\thefootnote}{}
\footnotetext{*First Author and Second Author contribute equally to this work.}
\footnotetext{\IEEEauthorrefmark{2}Corresponding Author.}
}

\IEEEdisplaynontitleabstractindextext

%
\IEEEpeerreviewmaketitle

\IEEEraisesectionheading{\section{Introduction}\label{sec:introduction}}
\IEEEPARstart{D}{epression}  is a common mental disorder characterized by a negative state of mind that persists for a long time. It constitutes a significant impairment to an individual's cognitive processes, behaviors, and emotional states~\cite{pampouchidou2017automatic}.
Furthermore, it poses a grave risk of precipitating self-harmful behaviors, including self-mutilation and suicidal tendencies~\cite{soloff1994self}. Timely diagnosis coupled with appropriate therapeutic interventions can alleviate of these distressing symptoms of patients. However, the diagnostic process is often laborious and relies on the expertise of psychologists. Investigating automatic depression detection methods can be imperative to enhance diagnostic efficiency.

With the rapid development of Deep Learning (DL), various DL-based methodologies for depression detection have been proposed~\cite{al2018detecting, bucur2023s, fan2024transformer, fang2023multimodal, mallol2019hierarchical, niu2020multimodal, rodrigues2019multimodal, xue2024fusing, zhang2024novel}, leveraging diverse data inputs such as speech, visual, and physiological signals.
Among these, speech data is relatively easier to obtain in many practical scenarios.
Thus, this paper focuses on the modeling methods using speech for depression detection.
Early depression detection methods~\cite{al2018detecting, lam2019context, xiao2021deep, ye2021multi} with speech data employ specialized pre-trained networks to extract depression-related acoustic and textual features,
followed by the application of various fusion strategies to integrate these features for detecting depression.
Later,
acknowledging that depression is marked by impaired emotion regulation~\cite{joormann2014emotion},
researchers have recognized the potential of additional emotional features derived from speech data for depression detection~\cite{qureshi2020improving, teng2024multi, wu2022climate, xuchangqing2024}.
They generally utilize the sentiment (emotion) label from the acoustic modality as the supervised information to derive auxiliary emotional features. These studies have demonstrated that the emotional features obtained from a single acoustic sentiment label are valuable for improving the diagnostic accuracy of depression.

In addition, patients with mental disorders, including those with depression, commonly exhibit special emotional expressions in their speech. For instance, individuals with depression might convey negative emotional content in an unexpectedly calm manner~\cite{wu2024Depression}, showing a high degree of inconsistency in emotional expressions. These distinctive patterns could serve as valuable indicators for symptom assessment. In related work on mood identification in bipolar disorder~\cite{niu2023capturing}, researchers have manually developed the Emotional MisMatch (EMM) features to capture the inconsistency between linguistic and acoustic emotional expressions, thereby enhancing the performance of the Hamilton Depression Rating Scale (HDRS) regression task. Paralleling this, the Emotion Context-Insensitivity (ECI) theory~\cite{rottenberg2005mood} also highlights that depressed patients exhibit a constrained range of emotional variation in their vocal expressions, irrespective of the conversational context. However, the phenomenon of emotional expression inconsistency has not been considered in prior depression detection studies.
This paper, therefore, attempts to utilize this inconsistency for speech-based depression diagnosis.

In this context, there are two challenges: effectively modeling the phenomenon of the emotional expression inconsistency and integrating it into current depression detection methods.
For the first challenge, we assume that the emotion in a given speech is consistent if both its acoustic and textual modalities share the same sentiment labels; otherwise, the emotion is inconsistent. Based on this assumption, a multimodal cross-attention method, leveraging consistent/inconsistent labels as supervision, is proposed to derive the inconsistency information from the feature sequences preprocessed by self-supervised learning (SSL) based models, referred to as Acoustic-Textual Emotional Inconsistency (ATEI).
This method firstly leverages a multihead self-attention mechanism to extract emotional expression patterns from both acoustic and textual domains, by analyzing the intricate temporal, local, and global dependencies related to emotional expressions. The cross-attention technique is then used to derive the degree of the mismatch between the emotional expressions within acoustic and textual data. By integrating this mismatch degree with the identified emotional expression patterns from both domains, we can effectively derive ATEI information that is highly relevant to depression.
For the second challenge, a Transformer-based framework is proposed to integrate the derived ATEI information with various fusion strategies for detecting depression. Moreover, we examined the relationship between the degree of inconsistency and individuals with different depression severities,
such as whether individuals experiencing more severe depression are more likely to utilize a monotonous tone when speaking sad textual content.
Thus, we introduced a learnable scaling factor that dynamically adjusts the ATEI information during the fusion process and explores the experiment to document that these feature representations are highly correlated with depression severity.

Beyond the methodology investigation, the selection and preparation of training data also warrant attention. The phenomenon of emotional expression inconsistency is frequently discernible within conversations that are replete with affective expression. The majority of existing public datasets, such as DAIC-WOZ~\cite{gratch2014distress} and EATD~\cite{shen2022automatic}, primarily consist of short responses to standardized questionnaires. In contrast, conversational data provides a more diverse and nuanced range of emotional expressions across both acoustic and textual modalities, offering a more comprehensive view of the patient's emotional state. Consequently, counseling conversational data is recognized as a valuable dataset for the detection of depression in this context.

The main contribution of this paper can be summarized as follows:
\begin{itemize}
\item[1.]{A multimodal cross-attention method, which uses distinct sentiment labels for different modalities, is proposed to effectively extract the ATEI information related to depression from the counseling conversational speech. The proposed ATEI extraction method is a data-driven approach that eliminates the need for manual intervention, and exhibits robust generalization capability.
  Compared to traditional emotional features derived from a single acoustic sentiment label, the proposed ATEI information using multiple sentiment labels may offer more nuanced insights into the mental state of
depressed patients, potentially enhancing the accuracy of depression diagnosis.
    }
\item[2.]{A Transformer-based framework using additional ATEI information is proposed for depression detection. To the best of our knowledge, this work is the first to incorporate emotional expression inconsistency information across different modalities into depression detection. Experimental results showed that the proposed depression detection system with additional ATEI information exhibited higher accuracy in comparison to the SOTA baseline without ATEI information. This suggests that inconsistency in emotional expression can serve as one of the important symptoms of depression.}
\item[3.]{This paper extensively investigates three different fusion strategies to integrate the acquired ATEI information into the mainstream depression detection methodologies. We enhance the fusion process by introducing a learnable scaling factor for the ATEI features. This adaptation allows the network to dynamically adjust the influence of ATEI information, facilitating the learning of representations that are more discriminative of depression severity. As a result, the final network becomes more adept at distinguishing between patients with varying degrees of depression, thereby significantly enhancing the precision of depression diagnosis.}
\end{itemize}

The rest of this paper is organized as follows. Section~\ref{sec:related-works} reviews existing works on automatic depression detection methods with acoustic and textual inputs. Section~\ref{sec:counseling-data} presents the data analysis between depressed patients and healthy controllers. The proposed framework using ATEI information for depression detection is introduced in Section~\ref{sec:proposed-framework}, followed by experiments and results in Section~\ref{sec:experiment}. Finally, conclusions and future works are drawn in Section~\ref{sec:conclusion}.

\section{Related works}
\label{sec:related-works}
Most depression detection methods concentrate on two aspects: feature extraction and model development. In this section, we will review DL-based approaches for automatic depression detection, with a focus on these two aspects. Furthermore, considering that counseling conversational includes both acoustic elements and their corresponding textual components, we will examine DL-based methods that either utilize the acoustic modality only or integrate both modalities for depression detection.

\subsection{Depression Detection Using Acoustic Modality}
\label{subsec:audio-only-ADD}
In early studies about feature extraction, most features are handcraft features~\cite{klatt1990analysis, france2000acoustical, alpert2001reflections}, including F0, loudness, speaking rate, jitter, shimmer, and others.
The use of handcraft features typically requires substantial expert involvement.
To alleviate the reliance on expert intervention, researchers use various deep neural networks to extract high-level semantic features~\cite{he2018automated, du2023depression, lam2019context, lu2022prediction, marriwala2023hybrid, niu2019automatic, ye2021multi, zhao2020hybrid}.
These networks have demonstrated greater reliability and efficiency in extracting depression-related features traditional manual feature extraction methods.
The commonly used neural network architectures include CNNs, LSTMs, and Transformers, among others.
However, these studies are constrained by the limitation of resources with labeled data, impeding their ability to capture comprehensive depression-related acoustic information~\cite{guo2023prompt}.

To address this issue, researchers have turned to transfer learning, which involves training a general model on large-scale, diverse data to be used in various related downstream tasks and domains.~\cite{bommasani2021opportunities}.
SSL is a popular method for pre-training foundation models.
The commonly used SSL-based models include Wav2Vec~\cite{baevski2020wav2vec}, HuBERT~\cite{hsu2021hubert}, Whisper~\cite{radford2023robust}, and WavLM~\cite{chen2022wavlm}.
By leveraging large amounts of unlabeled data, SSL-based models can develop a robust understanding of the inherent structure of speech and text.
Pre-training on tasks like predicting missing words or reconstructing masked segments of speech and text allows these models to capture a wide range of universal features related to depression, such as speaking style and emotional expressions.
For example, Huang et al.~\cite{huang2024research} fine-tuned the wav2vec2.0 model to encode audio, achieving excellent results on the DAIC-WOZ dataset.
Xue et al.~\cite{xue2024fusing} designed a multi-level feature interaction module to fuse the traditional acoustic features and wav2vec-derived deep embeddings, thereby obtaining a comprehensive audio-based deep emotional representation.
Wu et al.~\cite{wu2023self} leveraged wav2vec2.0, HuBERT, and WavLM to transfer emotion recognition knowledge to depression detection, surpassing mainstream methods on the DAIC-WOZ dataset.
Studies mentioned above highlight the advantage of SSL-based methods in enhancing depression detection with limited samples.



Apart from feature extraction, model construction is also a crucial part of depression detection. The mainstream models used for depression detection include CNNs, LSTMs, and Transformers.
For instance, the study in~\cite{huang2024research} achieved notable success in depression detection by leveraging LSTM with the self-attention mechanism.
In~\cite{yin2023depression}, the authors introduced a deep learning framework integrating parallel CNNs and Transformers, featuring three parallel processing streams to capture different aspects of local and temporal information, with its effectiveness validated through experiments on the DAIC-WOZ and MODMA datasets.

\subsection{Depression Detection Using Acoustic-Textual Modalities}
\label{subsec:audio-text-ADD}
Previous studies have shown that diverse modalities, encompassing both acoustic and textual features, exhibit complementary properties and possess the capability to enhance the performance of depression detection systems~\cite{al2018detecting,rohanian2019detecting}. Effective feature representations and appropriate fusion methods are crucial for multimodal depression detection.

For the audio modality, features commonly used in depression detection in recent years are the deep features derived from SSL-based models, as detailed in Section~\ref{subsec:audio-only-ADD}. In this paper, the SSL-based methods will be also used for processing the acoustic inputs.
For text modality, commonly used features in depression detection can be divided into two categories: statistical features and text-to-vector embeddings~\cite{muzammel2021end}.
Statistical features pertain to the quantitative analysis of spoken utterances by patients with depression during counseling, encompassing metrics such as the frequency of nouns, adjectives, pronouns, and words associated with social processes and psychological states.
Text-to-vector embedding is typically obtained via language models like Paragraph Vector (PV)~\cite{le2014distributed}, XLNet~\cite{yang2019xlnet}, BERT~\cite{devlin2018bert}, and their variants to represent each word using a vector. Previous studies have shown that the text-to-vector embeddings extracted from BERT and its variances contain lots of semantic information related to depression~\cite{zeberga2022retracted,bucur2023s}, thereby facilitating their effective utilization in enhancing the overall performance of depression detection systems. In this paper, BERT-based pre-trained models will be considered to be used for processing the textual data.

Furthermore, acknowledging that depression is characterized by dysregulated emotion~\cite{joormann2014emotion}, the potential of emotional features extracted from speech data for depression detection has been increasingly recognized~\cite{qureshi2020improving, teng2024multi, wu2022climate, xuchangqing2024}.
In these methodologies, the sentiment of the acoustic modality typically serves as the supervised label. The emotional state of depressed patients is described by using the hidden outputs generated by deep neural networks. These emotional features are then used as the auxiliary information and integrated with depression-related acoustic and textual features for depression detection.
Experimental results in~\cite{qureshi2020improving, teng2024multi, wu2022climate, xuchangqing2024} have shown that additional emotional features derived from a single acoustic sentiment label can enhance the diagnostic precision of depression detection systems.

Modality fusion techniques are also important for multimodal depression detection. The fusion techniques can be categorized into two major types: decision-level fusion and feature-level fusion.
Decision-level fusion is based on the decision from each modality~\cite{wu2022climate,saggu2022depressnet,fang2023multimodal}. It integrates the predictions derived from individual depression detection models, corresponding to each modality (encompassing probabilities, confidence scores, classification labels, or alternative prediction outcomes).
Compared to the decision-level fusion methods, feature-level fusion is the most commonly used strategy for multimodal systems~\cite{rodrigues2019multimodal, park2022design, fang2023multimodal, fan2024transformer} in recent years.
In the depression detection task, the features extracted from different modalities often exhibit complementarity. Leveraging feature-level fusion, these features are integrated into a unified feature vector through a prescribed fusion algorithm. This process can capture the interrelationships and complementary nature of the features across different modalities, thereby improving the prediction accuracy of the severity of depression~\cite{gratch2014distress,shen2022automatic}.
The commonly used feature-level fusion strategies include addition, multiplication, and concatenation.
This paper will delve into the integration of the proposed ATEI information into SOTA depression detection methods, focusing on these three fusion strategies.



\section{Counseling Conversational Data}
\label{sec:counseling-data}

The dataset used for depression detection comprises a collection of online telephone counseling sessions spanning 22.9 hours, encompassing a total of 272 individuals aged from 12 to 45.
This dataset comprises healthy control individuals with the Self-rating Depression Scale (SDS)~\cite{ZUNG1965} and Self-Rating Anxiety Scale (SAS)~\cite{ZUNG1971371}, as well as depressed subjects with the Hamilton Depression Rating Scale (HAMD).
The depressed patients within the dataset have undergone professional diagnosis by certified psychiatrists, and the healthy control individuals have never experienced depression or any other psychiatric illnesses.
More detailed information of the dataset is provided in TABLE~\ref{tab-dataset}.
\begin{figure*}[htb]
\centering
\includegraphics[width=7in]{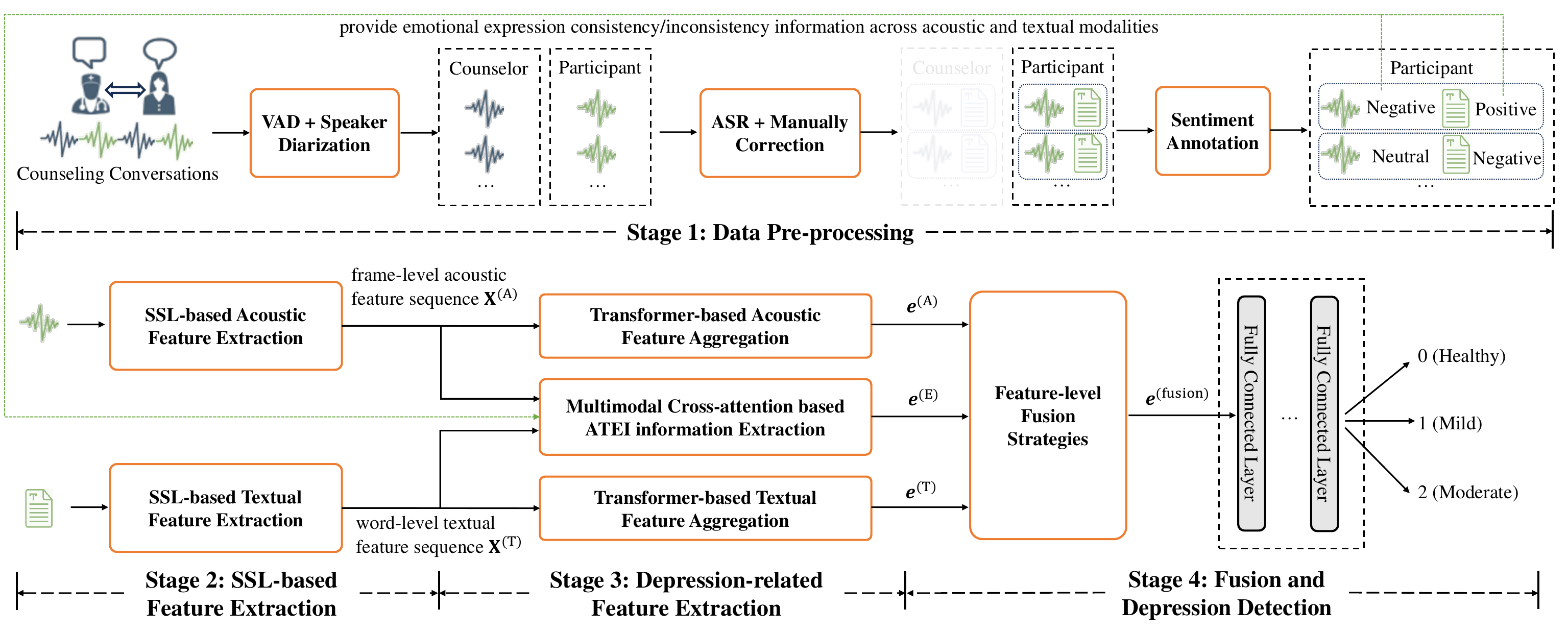}
\caption{Proposed Transformer-based framework using additional Acoustic-Textual Emotional Inconsistency (ATEI) information for automatic depression detection. Both $\mathbf{X}^{(\text{A})}$ and $\mathbf{X}^{(\text{T})}$ contain the short-term universal information related to depression. $\boldsymbol{e}^{(\text{A})}$, $\boldsymbol{e}^{(\text{T})}$ and $\boldsymbol{e}^{\text{(E)}}$ represent the acoustic, textual and ATEI features related to depression over a long time range, respectively.}
\label{fig:proposed-framework}
\end{figure*}

\begin{table}[htbp]
\caption{Details of the counseling conversational dataset}
\label{tab-dataset}
	\begin{center}
		{
          \begin{tabular}{l | c | c | c | c } \hline

                \textbf{Class} & \textbf{Scale Score} & \textbf{\#Male/Female}  & \textbf{\#Hours} & \textbf{\#Seg}
                \\
                \hline
                Healthy & 25-45 (SDS\&SAS) & 53/47  & 9.3 & 2700 \\
                \hline
                Mild & 8-13 (HAMD) & 49/51  & 8.5 & 3924 \\
                \hline
                Moderate & 14-18 (HAMD) & 36/36  & 5.1 & 2708 \\
                \hline
			\end{tabular}
        }
\end{center}
\end{table}

The counseling was administered by seasoned psychological counselors with expertise in their respective fields. During the counseling process, all psychological counselors will employ a variety of strategies~\cite{clark2014empathy}, such as empathy-driven techniques to establish trust,
in order to encourage participants to produce more voice content and openly express their genuine emotions.
The counseling sessions encompassed diverse conversational topics such as sleep, work, learning, and interpersonal relationships.
Each counseling session is accompanied by a corresponding single-channel WAV file, recorded at a sampling rate of 8kHz.
%

\section{Proposed Framework}
\label{sec:proposed-framework}

Drawing upon our preceding investigations~\cite{wu2024Depression}, depressed individuals exhibit inconsistent emotional expressions in the acoustic and textual modalities compared to healthy controls. This phenomenon, referred to as Acoustic-Textual Emotional Inconsistency (ATEI), holds the potential for distinguishing between healthy controls and depressed patients. Thus, the primary focus of this article is the development of data-driven methodologies aimed at effectively extracting ATEI information and its subsequent utilization in depression detection modeling.
Along this line, this paper proposes a novel Transformer-based framework using additional ATEI information for automatic depression detection. As shown in Fig.~\ref{fig:proposed-framework},
the proposed framework is structured into four stages.
The initial stage involves data preprocessing, with the goal of acquiring both acoustic segments and their corresponding transcribed textual segments. In the second stage, we employ SSL-based feature extraction methods, which enable the rapid acquisition of short-term universal feature sequences, containing information like speech speed, intonation, syntactic structure, textual semantics, and others. The third stage derives various depression-related features across a long time range from these universal feature sequences. Lastly, we integrate the acquired depression-related features through various feature-level fusion strategies, applying multiple fully connected layers to facilitate the prediction of depression severity.

\subsection{Data Pre-processing}
\label{subsec:data-processing}
Due to the significant presence of silence during each counseling, which can adversely affect the accuracy of depression detection,
we employed the open-source toolkit pyannote.audio\footnote{\url{https://github.com/pyannote/pyannote-audio}} to automatically segment each audio file. This automatic segmentation was followed by a manual review to ensure the precision of the segmentation process. In general, each acoustic segment data contained more than one sentence.
Furthermore, each acoustic segment was accompanied by a manually corrected transcription. The transcriptions were initially generated using the WeNet\footnote{\url{https://github.com/wenet-e2e/wenet}} toolkit and subsequently verified through manual double-checking.
To incorporate the inconsistency in emotional expressions between acoustic and textual modalities into automatic depression detection methods, sentiment annotation is essential. In this study, we assigned sentiment labels of positive, negative, and neutral to each audio and textual segment. For the sentiment annotation of the audio data, a single-blind approach was employed, ensuring that each annotator was unaware of the audio's origin. Concisely, each audio segment was labeled by five individuals, and the label with the highest consensus was assigned as the sentiment label for that segment. For the text data, a sentiment analysis method based on the HowNet lexicon~\cite{dong2003hownet} was utilized to derive the sentiment labels.
If the aggregated sentiment score of a text segment was greater than zero, it was classified as positive; if less than zero, it was deemed negative. In cases where the total sentiment score of a text segment was zero, the sentiment label was assigned as neutral.

\subsection{SSL-based Feature Extraction}
\label{subsec:feature-extraction}
%
Pre-trained models leveraging SSL techniques have garnered remarkable achievements. These models exhibit a remarkable ability to extract intricate features from unlabeled data to address data scarcity.
Their robust generalization and transfer learning capabilities have made them invaluable across multiple fields, such as depression detection.
Our objective in this section is to leverage the deep data representations learned by these models to facilitate the easier acquisition of robust ATEI information in subsequent processes.

For the speech modality, commonly used pre-trained models for the feature extraction include Wav2Vec~\cite{baevski2020wav2vec}, HuBERT~\cite{hsu2021hubert}, Whisper~\cite{radford2023robust}, and WavLM~\cite{chen2022wavlm}.
In this paper, we investigated these SSL-based models for acoustic feature extraction.
Throughout the entire training process, the SSL-based model parameters remained fixed.
Additionally, since an abundance of emotional expression information is contained in the intermediate layer outputs of SSL-based models~\cite{chen2023exploring, gat2022speaker, morais2022speech}, we have chosen the outputs from the 12th block of Wav2Vec, Whisper, WavLM, and HuBERT models as the inputs for further modeling in this paper.
The acoustic features derived from these SSL-based models are at frame-level granularity.

For the text modality, previous studies have demonstrated that employing BERT~\cite{devlin2018bert} can grasp long-range linguistic dependencies through a multi-layered Transformer structure, thus enabling the acquisition of robust textual features for each word. These word-level textual features encompass substantial semantic information that can be harnessed effectively for depression detection. RoBERTa ~\cite{adoma2020comparative, liu2019roberta}, a pre-trained language model refined from the original BERT architecture, achieves superior performance compared to the latter by adjusting critical hyperparameters and training techniques. RoBERTa has demonstrated remarkable performance in downstream applications such as sentiment analysis, text classification, and question answering systems.
In this paper, we investigated both BERT and RoBERTa for textual feature extraction.

\subsection{Depression-related Feature Extraction}
\label{subsec:depression-related-feats-extraction}

\subsubsection{\textbf{Acoustic and textual features related to depression}}
\begin{figure}[ht]
    \centering
    \includegraphics[width=3.5in]{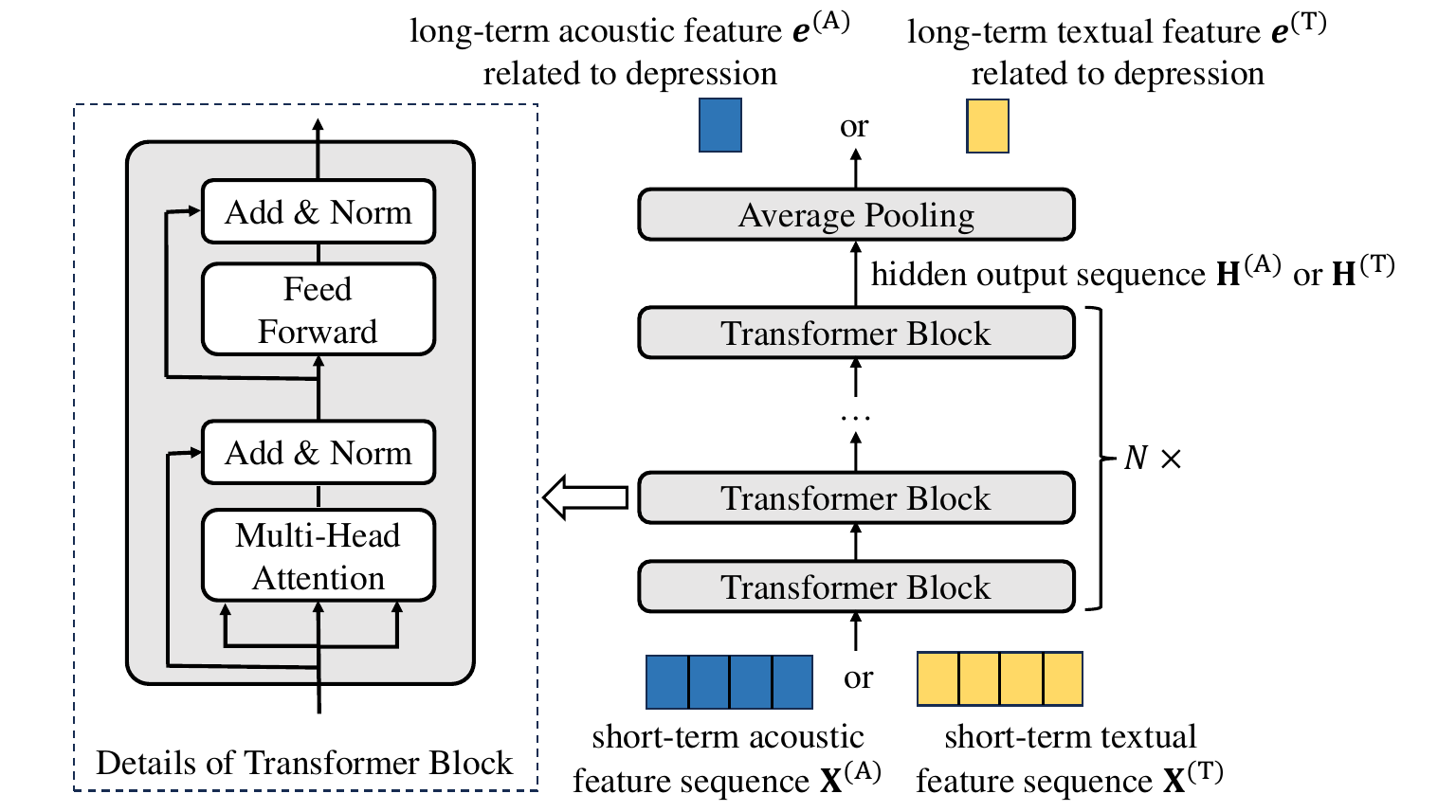}
    \caption{Transformer-based feature aggregation for extracting the acoustic and textual features related to depression.}
    \label{fig:feature-aggregation}
\end{figure}
Utilizing the SSL-based feature extraction method described in Section~\ref{subsec:feature-extraction}, we can derive frame-level acoustic features and word-level textual features.
These short-term features mainly contain local information in speech or transcriptions, such as the acoustic properties of individual speech frames or the semantic content of individual words. However, detecting depression often requires comprehensive consideration of a wider range of contextual information, including changes in tone of speech, overall emotional tendencies of sentences in the text, and other things. Therefore, using short-term features alone often cannot fully reflect the overall state of patients with depression. We need to use feature aggregation methods to obtain long-term depression-related features from the acquired short-term SSL-based feature sequences.

Common feature aggregation approaches include both traditional statistical methods like average pooling, as well as modern deep learning-based techniques. The latter employs diverse network architectures, such as convolutional neural networks (CNN), recurrent neural networks (RNN), and Transformers, to efficiently aggregate short-term features across the temporal dimension. Compared to traditional RNN and CNN, the Transformer architecture excels in capturing long-range dependencies, coupled with its outstanding performance in speech and natural language processing (NLP) tasks. Consequently, this paper leveraged the Transformer for aggregating short-term features across both acoustic and textual modalities.

As shown in Fig.~\ref{fig:feature-aggregation}, assuming that $\mathbf{X}^{(\text{A})}$ and $\mathbf{X}^{(\text{T})}$ are the acquired frame-level acoustic feature sequence and word-level textual feature sequence, respectively.
\begin{equation}
  \begin{split}
  \mathbf{X}^{(\text{A})} &= \left[\boldsymbol{x}^{(\text{A})}_{1}, \boldsymbol{x}^{(\text{A})}_{2}, \ldots, \boldsymbol{x}^{(\text{A})}_{ T_1}\right]\in\mathbb{R}^{T_1\times D}, \\
  \mathbf{X}^{(\text{T})} &= \left[\boldsymbol{x}^{(\text{T})}_{1}, \boldsymbol{x}^{(\text{T})}_{2}, \ldots, \boldsymbol{x}^{(\text{T})}_{T_2}\right]\in\mathbb{R}^{T_2\times D}
  \end{split}
\end{equation}
where $\boldsymbol{x}^{(\text{A})}_{i}$ is the $D$-dimensional acoustic feature vector of $\mathit{i}$-th frame , $\boldsymbol{x}^{(\text{T})}_{j}$ is the $D$-dimensional textual feature vector of $\mathit{j}$-token, $T_1$ is the number of frames of current acoustic segment and $T_2$ is the number of tokens of current textual segment.
Subsequently, $\mathbf{X}^{(\text{A})}$ and $\mathbf{X}^{(\text{T})}$ are individually processed through dedicated neural network architectures, each comprising several Transformer blocks. This facilitates the integration of short-term features into segment-level representations. Each Transformer block includes:

(a) Multi-Head Attention: The multi-head attention mechanism splits the inputs into multiple subspaces and applies single-head attention in each subspace. Assuming the input of current Transformer block is $\mathbf{X}$ and there are $h$ heads, for each head $\mathit{i}$, we have:
\begin{equation}
\begin{split}
\boldsymbol{Q}_i &= \mathbf{X} \boldsymbol{W}^{Q}_i, \\
\boldsymbol{K}_i &= \mathbf{X} \boldsymbol{W}^{K}_i, \\
\boldsymbol{V}_i &= \mathbf{X} \boldsymbol{W}^{V}_i
\end{split}
\end{equation}
\begin{equation}
\begin{split}
\mathrm{head}_i &= \mathrm{Attention}(\boldsymbol{Q}_i, \boldsymbol{K}_i, \boldsymbol{V}_i) \\
&= \mathrm{softmax} \left( \frac{\boldsymbol{Q}_i (\boldsymbol{K}_i)^\top}{\sqrt{d_k}} \right) \boldsymbol{V}_i
\end{split}
\label{eq:head-attention}
\end{equation}
\begin{equation}
\begin{split}
\mathbf{Z} &= \mathrm{MultiHead}(\boldsymbol{Q}_i, \boldsymbol{K}_i, \boldsymbol{V}_i) \\
&= \mathrm{Concat}(\mathrm{head}_1, \mathrm{head}_2, \ldots, \mathrm{head}_h) \boldsymbol{W}^O
\end{split}
\end{equation}
where $\boldsymbol{Q}_i, \boldsymbol{K}_i, \boldsymbol{V}_i$ are the query, key, and value matrices of the $\mathit{i}$-th head, respectively;
$\boldsymbol{W}^{Q}_i, \boldsymbol{W}^{K}_i, \boldsymbol{W}^{V}_i$ are the learnable parameter matrices that project the input $\mathbf{X}$ into the required dimensional space;
${d_k}$ is the scaling factor; $\boldsymbol{W}^O$ is used to project the concatenated multi-head attention output back to the original dimension $D$.

(b) Feed Forward: After the output from each attention head, the information is further processed through a feed-forward network consisting of a linear layer with ReLU activation function. The formula is defined as:
\begin{equation}
\mathrm{FFN}(\mathbf{Z})=\mathrm{ReLU}(\mathbf{Z}\boldsymbol{W}_1+\boldsymbol{b}_1)\boldsymbol{W}_2+\boldsymbol{b}_2
\end{equation}
where $\boldsymbol{W}_{1}$ and $\boldsymbol{W}_{2}$ are the weights of the linear layers, $\boldsymbol{b}_{1}$ and $\boldsymbol{b}_{2}$ are the bias terms.

(c) Add \& Norm: The output of each sub-layer will be added to the original input, and then layer normalization is performed, which is shown in the left of Fig.~\ref{fig:feature-aggregation}.

After traversing $N$ blocks of the Transformer Encoder depicted in Fig.~\ref{fig:feature-aggregation}, we are able to obtain $\mathbf{H}^{(\text{A})}$ and $\mathbf{H}^{(\text{T})}$ from $\mathbf{X}^{(\text{A})}$ and $\mathbf{X}^{(\text{T})}$, respectively. They are defined as:
\begin{equation}
\begin{split}
\mathbf{H}^{(\text{A})}=[\boldsymbol{h}^{(\text{A})}_{1},\boldsymbol{h}^{(\text{A})}_{2},...,\boldsymbol{h}^{(\text{A})}_{T_1}]\in\mathbb{R}^{T_1\times D},\\
\mathbf{H}^{(\text{T})}=[\boldsymbol{h}^{(\text{T})}_{1},\boldsymbol{h}^{(\text{T})}_{2},...,\boldsymbol{h}^{(\text{T})}_{T_2}]\in\mathbb{R}^{{T_2}\times D}
\end{split}
\end{equation}
Subsequently, $\mathbf{H}^{(\text{A})}$ and $\mathbf{H}^{(\text{T})}$ are fed into an average pooling layer to derive segment-level embeddings $\boldsymbol{e}^{(\text{A})}$ and $\boldsymbol{e}^{(\text{T})}$, respectively. These embeddings will serve as depression-related features in the acoustic and textual domains in this paper. They are defined as:
\begin{equation}
\begin{split}
\boldsymbol{e}^{(\text{A})}=\mathrm{Avg}(\mathbf{H}^{(\text{A})})=\frac{1}{T_1}\sum_{i=1}^{T_1}\boldsymbol{h}^{(\text{A})}_{i}\in\mathbb{R}^{1\times D},\\
\boldsymbol{e}^{(\text{T})}=\mathrm{Avg}(\mathbf{H}^{(\text{T})})=\frac{1}{T_2}\sum_{j=1}^{T_2}\boldsymbol{h}^{(\text{T})}_{j}\in\mathbb{R}^{1\times D}
\end{split}
\end{equation}

\subsubsection{\textbf{ATEI information}}
\label{subsubsec:ATEI-depression-related-feats-extraction}
\begin{figure}[ht]
    \centering
    \includegraphics[width=3.5in]{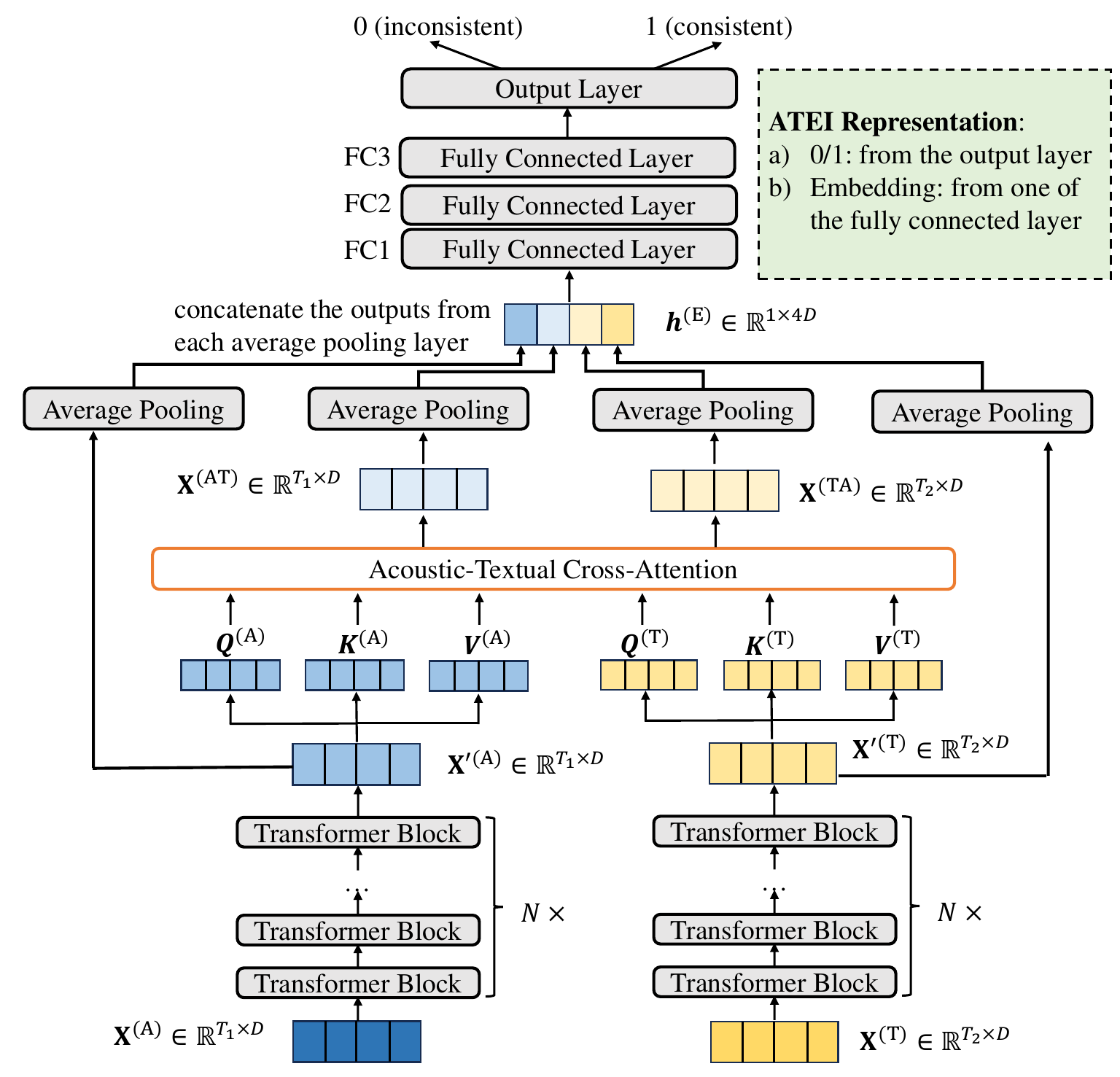}
    \caption{Multimodal cross-attention method for extracting ATEI information.}
    \label{fig:ATEI-cues-extraction}
\end{figure}



The proposed multimodal cross-attention model for ATEI information extraction is shown in Fig.~\ref{fig:ATEI-cues-extraction}.
The supervised labels of this model are consistent (labeled as ``1'') and inconsistent (labeled as ``0''). Precisely, when an acoustic segment and its corresponding textual segment share the same sentiment label, the supervised label for this acoustic-textual segment pair is set to ``1''. Conversely, when the sentiment labels between the acoustic and textual segments do not match, it is assigned ``0''.

As previously mentioned, the inputs $\mathbf{X}^{(\text{A})}$ and $\mathbf{X}^{(\text{T})}$ in Fig.~\ref{fig:ATEI-cues-extraction} contain universal information like speech speed, intonation, syntactic
structure, textual semantics, and others. This information is closely associated with emotional expression.
After passing through $N$ Transformer blocks, we can obtain the intricate local and global temporal dependencies related to emotional expressions in the acoustic and textual domains, respectively, denoted by $\mathbf{X}^{\prime(\text{A})}\in\mathbb{R}^{T_1\times D}$ and $\mathbf{X}^{\prime(\text{T})}\in\mathbb{R}^{T_2\times D}$.
And then $\mathbf{X}^{\prime(\text{A})}$ undergoes a linear transformation to yield $\boldsymbol{Q}^{(\text{A})}$, $\boldsymbol{K}^{(\text{A})}$, $\boldsymbol{V}^{(\text{A})}$. Similarly, $\mathbf{X}^{\prime(\text{T})}$ is transformed to produce $\boldsymbol{Q}^{(\text{T})}$, $\boldsymbol{K}^{(\text{T})}$, $\boldsymbol{V}^{(\text{T})}$. Through the acoustic-textual cross-attention layer, $\mathbf{X}^{(\text{AT})}$ and $\mathbf{X}^{(\text{TA})}$ are obtained by:
\begin{equation}
\mathbf{X}^{(\text{AT})}=\mathrm{softmax}\left(\frac{\boldsymbol{Q}^{(\text{A})}\boldsymbol{K}^{(\text{T})^\top}}{\sqrt{d_k}}\right)\boldsymbol{V}^{(\text{T})}\in\mathbb{R}^{T_1\times D}
\label{eq:multimodal-cross-att-a}
\end{equation}
\begin{equation}
\mathbf{X}^{(\text{TA})}=\mathrm{softmax}\left(\frac{\boldsymbol{Q}^{(\text{T})}\boldsymbol{K}^{(\text{A})^\top}}{\sqrt{d_k}}\right)\boldsymbol{V}^{(\text{A})}\in\mathbb{R}^{T_2\times D}
\label{eq:multimodal-cross-att-b}
\end{equation}


Referring to the consistent or inconsistent supervised labels depicted in Fig.~\ref{fig:ATEI-cues-extraction} and as indicated by Equation~(\ref{eq:multimodal-cross-att-a}), the cross-attention score $\mathrm{softmax}(\cdot)$ is supposed to measure the degree of correlation between textual and acoustic emotional expressions. In other words, it is used to represent the mismatch level between each word and its corresponding acoustic sentiment. In addition, $\mathbf{X}^{(\text{AT})}$  is calculated from the feature $\boldsymbol{V}^{(\text{T})}$, which contains emotional expression information in the textual domain. This implies that the derived feature $\mathbf{X}^{(\text{AT})}$  represents the abnormal emotional expression patterns of depressed patients observed in the textual domain.
Similarly,  $\mathbf{X}^{(\text{TA})}$ in Equation~(\ref{eq:multimodal-cross-att-b}) represents the abnormal emotional expression patterns of depressed patients observed in the acoustic domain.
$\mathbf{X}^{\prime(\text{A})}$, $\mathbf{X}^{\prime(\text{T})}$, $\mathbf{X}^{(\text{AT})}$, and $\mathbf{X}^{(\text{TA})}$ are each processed by an average pooling layer and then concatenated to form the vector $\boldsymbol{h}^{\text{(E)}}\in\mathbb{R}^{1\times 4D}$:
\begin{equation}
\boldsymbol{h}^{\text{(E)}}=
[\mathrm{Avg}(\mathbf{X}^{\prime(\text{A})});\mathrm{Avg}(\mathbf{X}^{(\text{AT})});\mathrm{ Avg}(\mathbf{X}^{(\text{TA})});\mathrm{ Avg}(\mathbf{X}^{\prime(\text{T})})]
\label{eq:ATEI-hidden}
\end{equation}
According to Equation~(\ref{eq:ATEI-hidden}), $\boldsymbol{h}^{\text{(E)}}$ should contain the abnormal emotional expression patterns in both acoustic and textual domains.
Finally, the concatenated vector $\boldsymbol{h}^{\text{(E)}}$ is fed into three fully connected layers (FC1, FC2, and FC3) for distinguishing whether the current emotional expression across acoustic and textual modalities is consistent.


As discussed before, the representation of ATEI information is important for depression detection.
A straightforward approach for describing the ATEI information is directly utilizing the ``0/1'' output generated by the multimodal cross-attention network shown in Fig.~\ref{fig:ATEI-cues-extraction}.
In addition, we can also use the probability distribution of ``0/1'' generated from the output layer without softmax as the representation of ATEI information.
This ``0/1'' representation methodology can provide a direct reflection of the degree of emotional alignment between audio and textual components.

Beyond the straightforward ``0/1'' representation, an alternative approach is to utilize the hidden outputs within the multimodal cross-attention network to represent ATEI information, commonly referred to as ``Embedding'' representation. Given that $\boldsymbol{h}^{\text{(E)}}$ contains the abnormal emotional expression patterns in both acoustic and textual domains, the acquired embedded features inherently capture essential information for distinguishing the alignment of emotional expressions across these domains. This capability holds significant value for downstream tasks, particularly within the realm of depression detection.
In this paper, we explore the outputs from each fully connected layer (FC1, FC2, and FC3), as shown in Fig.~\ref{fig:ATEI-cues-extraction}, for the purpose of representing ATEI information, denoted by $\boldsymbol{e}^{\text{(E)}}$,
\begin{equation}
    \boldsymbol{e}^{\text{(E)}}=\mathrm{FC}(\boldsymbol{h}^{\text{(E)}})
    \label{eq:ATEI-ori}
\end{equation}
where $\mathrm{FC}$ denotes the operator of the fully connected layers.

In the preliminary experiments, we have tried to directly leverage ATEI information for depression detection. However, we noted that the depression detection system only using the ATEI embeddings as inputs gave an accuracy of approximately 40\%. This suggests that, although ATEI information encompasses distinguishing features between healthy individuals and depressed patients, it may not be adequate for a direct depression diagnosis or as the sole diagnostic criterion. Instead, ATEI information can serve as an auxiliary feature in the depression diagnosis process, where a thorough evaluation necessitates a comprehensive assessment of the patient's behavioral symptoms, psychological state, and other pertinent data extracted from counseling sessions.
Consequently, it is imperative to conduct further exploration into the fusion strategies for integrating ATEI information with the depression-related acoustic and textual features outlined before.

\subsection{Fusion Strategies}
\label{subsec:fusion}
This paper investigates three distinct feature-level fusion strategies for the acquired acoustic and textual features associated with depression, as well as the ATEI information. These fusion strategies include addition, multiplication, and concatenation. They are defined as:

a) \textit{Addition}:
\begin{equation}
  \boldsymbol{e}_{\text {fusion }}=\boldsymbol{e}^{(\text{A})} \oplus \boldsymbol{e}^{(\text{T})} \oplus \boldsymbol{e}^{\text{(E)}}
\end{equation}
where $\oplus$ represents the element-wise addition.

b) \textit{Multiplication}:
\begin{equation}
    \boldsymbol{e}_{\text {fusion }}=\boldsymbol{e}^{(\text{A})} \odot \boldsymbol{e}^{(\text{T})}  \odot \boldsymbol{e}^{\text{(E)}}
\end{equation}
where $\odot$ represents the element-wise multiplication.

c) \textit{Concatenation}:
\begin{equation}
    \boldsymbol{e}_{\text {fusion }}=[\boldsymbol{e}^{(\text{A})}, \boldsymbol{e}^{(\text{T})}, \boldsymbol{e}^{\text{(E)}}]
\end{equation}

Furthermore, we found that the degree of the ATEI differs among individuals with different levels of depression severity. For instance, individuals suffering from more severe depression would prefer to employ a flat tone to express sad text content.
In light of this observation, a learnable scaling factor $\boldsymbol{\alpha}=[\alpha_1,\alpha_2,...,\alpha_D]$ is introduced to dynamically modulate the ATEI.
This approach aims to precisely capture the correlation between the degree of ATEI information and the severity of depression, thereby enhancing the model's capacity to derive feature representations that robustly correlate with the severity of depressive symptoms, and ultimately improving the precision of diagnostic detection.
Thus, Equation~(\ref{eq:ATEI-ori}) can be revised as follows:
\begin{equation}
    \boldsymbol{e}^{\text{(E)}}=\boldsymbol{\alpha} \odot \mathrm{FC}(\boldsymbol{h}^{\text{(E)}}),
    \sum_{i=1}^{D}\alpha_i=1
    \label{eq:ATEI-new}
\end{equation}
where $\odot$ represents the element-wise multiplication.

It is noteworthy that when the ATEI information is encoded using the ``Embedding'' approach, $\boldsymbol{e}^{\text{(E)}}$ can be obtained from Equation~(\ref{eq:ATEI-ori}) or (\ref{eq:ATEI-new}).
Specifically, when the ATEI information is encoded using the ``0/1'' approach, only feature concatenation is considered, as the dimensions among $\boldsymbol{e}^{\text{(E)}}\in\mathbb{R}^{1\times 1}$, $\boldsymbol{e}^{(\text{A})}\in\mathbb{R}^{1\times D}$, and $\boldsymbol{e}^{(\text{T})}\in\mathbb{R}^{1\times D}$ are different.
After the fusion process, multiple fully-connected layers are utilized to predict the severity of depression, which is shown in Fig.~\ref{fig:proposed-framework}.
Moreover, to ensure robust training of the proposed model, this study employs an incremental training strategy for optimizing the model.
Initially, the multimodal cross-attention neural network for extracting ATEI information undergoes pre-training using the cross-entropy $\mathcal{L}_{\text {ATEI}}$, followed by joint training with the entire network. The loss function for the joint training stage is formulated as follows:
\begin{equation}\mathcal{L}_{\text {Total }}=\mathcal{L}_{\text {Depression }}+\mathcal{L}_{\text {ATEI}}\end{equation}
where $\mathcal{L}_{\text {Depression}}$ is the cross-entropy for optimizing the depression detection task, and $\mathcal{L}_{\text {ATEI}}$ is the cross-entropy for optimizing the emotion consistency recognition task.

\begin{table*}[htbp]
\renewcommand{\arraystretch}{1.2}
\caption{Performance of various baseline systems using depression-related Acoustic (A) and Acoustic-Textual (A+T) features.}
\begin{center}
    \scalebox{1}[1]  {
       \begin{tabular}{c||c|c|c|cccc|cccc } \hline
       \specialrule{1.5pt}{0pt}{0pt} 
            \multirow{2}{*}{Features} & \multicolumn{2}{c|}{SSL-based Methods}& Acoustic-Textual & \multicolumn{4}{c|}{Segment-level} & \multicolumn{4}{c}{Subject-level} \\ \cline{2-3}\cline{5-12}
             & Acoustic & Textual & Fusion Strategies & Acc & F1 & Pre & Rec & Acc & F1 & Pre & Rec \\ \hline
             \multirow{4}{*}{A} &Wav2Vec& $\times$& $\times$ & 56.24 & 57.37 & 56.19 & 58.47 & 68.01 & 68.91 & 68.01 & 69.83 \\
             & Whisper& $\times$ & $\times$& 56.95 & 57.83 & 56.81 & 58.88 &  71.32 & 70.38 & 69.87 & 70.89\\
             & WavLM & $\times$ & $\times$& 56.24 & 57.41 & 56.35 & 58.51 &  71.69 & 72.08 & 71.77 & 72.39\\
             &Hubert& $\times$ & $\times$&60.22 & 60.94 & 60.33 & 61.56 & \textbf{72.79} & \textbf{72.43} & \textbf{72.13} & \textbf{72.74} \\
             \hline\hline
             \multirow{6}{*}{A+T}  &\multirow{6}{*}{Hubert}&\multirow{3}{*}{BERT} & Add & 61.64 & 63.10 & 62.14 & 64.09 & 74.26 & 73.86 & 73.38 & 74.33 \\
             &&& Mult & 61.85 & 62.83 & 61.69 & 64.01& 74.63 & 74.68 & 74.29 & 75.07 \\
             &&& Concat & 62.32 & 63.35 & 62.23 & 64.52  & 75.37 & 75.22 & 74.87 & 75.57  \\\cline{3-12}

             & & \multirow{3}{*}{RoBERTa} & Add & 62.83 & 63.78 & 63.03 & 64.54 & 74.26 & 74.06 & 73.67 & 74.46 \\
             &&& Mult &  63.62 & 64.28 & 63.61 & 64.97& 75.00 & 75.21 & 74.65 & \textbf{75.78} \\
             &&& Concat & 64.16 & 64.81 & 64.10 & 65.53 & \textbf{75.74} & \textbf{75.31} & \textbf{74.95} & 75.67 \\
       \hline
       \specialrule{1.5pt}{0pt}{0pt} 
       \end{tabular}    }
\label{tab:baseline}
\end{center}
\end{table*}

\section{Experiments}
\label{sec:experiment}
Experiments were conducted on the dataset described in Section~\ref{sec:counseling-data}. The criteria used to evaluate the performance of the proposed framework include accuracy, precision, recall, and the F1 score. For each model, a five-fold cross-validation approach was employed, wherein 80\% of the speaker data (constituting the training set) was utilized for model training, while the remaining 20\% (the testing set) was reserved for evaluation purposes. Within each fold, the data from the same speaker were exclusively assigned to either the training set or the testing set, ensuring their non-overlap.

\subsection{Experimental Setup}
\label{subsec:experiment-setup}

To better assess the proposed framework, we firstly investigated the SOTA neural network architectures utilized for the detection of depressive disorders in this section.
Utilizing the established SOTA neural network architectures for depression detection as a foundation, we validated the effectiveness of ATEI information, which was derived from the multimodal cross-attention method, in enhancing the accuracy of depression detection.
Furthermore, we investigated the most suitable methodologies for integrating ATEI into the SOTA neural network architectures designed for depression detection.
The acoustic-textual baseline employs a network architecture featuring parallel acoustic and textual branches, mirroring the design depicted in Fig.~\ref{fig:proposed-framework}, albeit with a notable exclusion: the multimodal cross-attention neural network for extracting ATEI information. The acoustic baseline has the same structure as the acoustic processing branch of the acoustic-textual baseline.

In all depression detection models, we employed a Transformer architecture comprising 12 encoder blocks, each equipped with an eight-headed attention mechanism.
The scaling factor ${d_k}$ of Equation~(\ref{eq:head-attention}) was set to 128. The average pooling approach was used to condense the features extracted by the Transformer blocks into a standardized dimensionality of 1024. In the depression detection stage, three fully connected layers, each comprising 1024 nodes, were used to predict the severity level of depression.

All models were implemented using PyTorch and trained with four NVIDIA RTX 4090 GPUs. The Adam optimizer was initialized with a learning rate of 0.00001, a batch size of 64 was adopted, and the training process was terminated after a maximum of 30 epochs. To derive the classification outcomes across multiple segments for each subject, we adopted a majority voting approach. In this approach, the definitive classification label for a subject was determined by the prevailing vote among the labels assigned to the individual segments, ensuring a robust aggregation of segment-level classifications into a unified subject-level classification.

\subsection{Results}
\label{subsec:result}
\subsubsection{\textbf{SOTA depression detection baseline}}

We first explored how to use counseling conversational data to build SOTA depression detection baseline systems.
The performance of the depression detection systems with various settings is shown in TABLE~\ref{tab:baseline}.
This table is divided into two part: the first part presents the ``A'' depression detection systems utilizing depression-related acoustic features, while the second are the ``A+T'' systems that incorporate both depression-related acoustic and textual features.
It can be observed from the first part of TABLE~\ref{tab:baseline} that the ``A'' depression detection system utilizing HuBERT outperformed the other three SSL-based models, achieving the subject-level accuracy rate of 72.79\%. This indicates that HuBERT can provide more depression-related acoustic information than the other models.
Furthermore, we compared the performance of the ``A+T'' depression detection systems that utilized RoBERTa and BERT for preprocessing textual inputs. Our findings indicate that the ``A+T'' systems based on RoBERTa slightly outperformed those using the BERT model.
It is also observed that the ``A+T'' baseline system outperformed its counterparts trained with only acoustic features.
This indicates that there is a complementarity between the deep acoustic and textual features. Specifically, the SOTA ``A+T'' baseline system, which employs HuBERT for acoustic feature extraction and RoBERTa for textual feature extraction, and utilizes a concatenation fusion strategy (illustrated in the last line of Table~\ref{tab:baseline}), achieved the best performance, with a subject-level accuracy of 75.74\%.

\begin{table*}[htbp]
\renewcommand{\arraystretch}{1.2}
\caption{Performance of various depression detection systems using additional emotional (E) features---ATEI. The ATEI information can be represented by the ``0/1'' or ``Embedding'' approaches described in Section~\ref{subsubsec:ATEI-depression-related-feats-extraction}. The ATEI embedding features are derived from different fully connected layer (FC1, FC2, and FC3) of the multimodal cross-attention network in Fig.~\ref{fig:ATEI-cues-extraction}, The concatenation fusion strategy is used for integrating the acoustic, textual, and ATEI features.}
\begin{center}
    \scalebox{1}[1]  {
    \begin{threeparttable}
       \begin{tabular}{c||c|c|cccc|cccc } \hline
       \specialrule{1.5pt}{0pt}{0pt} 
            \multirow{2}{*}{Features} & \multirow{2}{*}{ATEI Information} & \multirow{2}{*}{Layer} & \multicolumn{4}{c|}{Segment-level} & \multicolumn{4}{c}{Subject-level} \\ \cline{4-11}
             &&  & Acc & F1 & Pre & Rec & Acc & F1 & Pre & Rec \\ \hline
            A & $\times$& $\times$ & 60.22 & 60.94 & 60.33 & 61.56 & 72.79 & 72.43 & 72.13 & 72.74 \\\hline
              \multirow{5}{*}{A+E} & 0/1 & \multirow{2}{*}{$\times$} & 64.55 & 65.41 & 64.6 & 66.24 & 74.63 & 74.61 & 74.17 & 75.06 \\
               & 0/1(Probability) && 64.49 & 65.35 & 64.50 & 66.23 & 74.63 & 74.50 & 74.08 & 74.93 \\ \cline{2-11}
               & \multirow{3}{*}{Embedding} & FC1 & 64.80 & 65.25 & 64.92 & 65.59 & \textbf{76.84} & 76.42 & 76.30 & 76.54 \\
               & & FC2 & 65.38 & 65.92 & 65.31 & 66.55 & \textbf{76.84} & \textbf{76.80} & \textbf{76.67} & \textbf{76.93} \\
               & & FC3 & 64.02 & 66.46 & 66.07 & 66.86 & 76.10 & 75.69 & 75.77 & 75.61 \\
               \hline\hline
             A+T & $\times$& $\times$ & 63.62 & 64.28 & 63.61 & 64.97 & 75.74 & 75.31 & 74.95 & 75.67 \\ \hline
             A+T+E~\cite{xuchangqing2024}\tnote{*}  & $\times$ & $\times$ & 63.62 & 64.28 & 63.61 & 64.97 & 76.84 & 77.00 & 76.57 & 77.44 \\ \hline
              \multirow{5}{*}{A+T+E}  & 0/1 & \multirow{2}{*}{$\times$} & 65.93 & 66.59 & 65.85 & 67.34 & 76.84 & 76.41 & 76.15 & 76.67 \\
               & 0/1(Probability) && 65.53 & 66.08 & 65.48 & 66.69 & 76.47 & 75.9 & 75.73 & 76.07 \\ \cline{2-11}
               & \multirow{3}{*}{Embedding} & FC1 & 66.07 & 66.51 & 66.06 & 66.96  & 78.68 & 78.24 & 78.41 & 78.07 \\
               & & FC2 & 66.10 & 66.57 & 66.08 & 67.07& \textbf{79.41} & \textbf{78.90} & \textbf{79.06} & \textbf{78.74} \\
               & & FC3 & 65.76 & 66.3 & 65.71 & 66.9  & 78.31 & 77.94 & 78.02 & 77.87 \\
       \hline
       \specialrule{1.5pt}{0pt}{0pt} 
       \end{tabular}
       \begin{tablenotes}    
        \footnotesize               
        \item[*] This system was reimplemented as described in~\cite{xuchangqing2024}, where the emotional features were extracted from the hidden outputs of a deep neural network. The supervised information is the single acoustic sentiment label.
       \end{tablenotes}
    \end{threeparttable}
       }
\label{tab:integration-ATEI}
\end{center}
\end{table*}

\subsubsection{\textbf{Multimodal cross-attention method for extracting ATEI information}}

As discussed before, it has been observed that, unlike healthy individuals, depressed patients often exhibit inconsistency in emotional expression across different modalities.
These observations have the potential to enhance the accuracy of prevalent depression detection technologies.
In this section, we investigated the effectiveness of extracting ATEI information related to depression using the multimodal cross-attention method described in Section~\ref{subsubsec:ATEI-depression-related-feats-extraction}.
The performance of the depression detection systems using additional ATEI information is presented in TABLE~\ref{tab:integration-ATEI}.
From the results in TABLE~\ref{tab:integration-ATEI}, we found five major trends.
\begin{itemize}
\item[I)]{
The depression detection systems that incorporated additional ATEI information demonstrated higher accuracy compared to their corresponding baseline systems lacking ATEI information, with an absolute improvement in subject-level accuracy ranging from 0.73\%-4.05\%.
This indicates that the proposed multimodal cross-attention method can effectively extract depression-related ATEI information from counseling conversational data. Additionally, the ATEI features can offer complementary information to the original acoustic and textual features associated with depression, thereby enhancing the model's performance  in a more comprehensive  fashion.}
\item[II)]{It is interesting that simply introducing additional ATEI information represented by ``0/1'' can also improve the system performance.
    This enhancement is observed both when utilizing the outputs from the final output layer with softmax and when utilizing the probability distributions from the output layer without softmax.
    Additionally, we found that the performance of the depression detection systems is similar whether using direct outputs or probability distribution outputs. This implies that, within the constraints of limited representation dimensions, employing more complex probability distribution representation methods can not provide more inconsistent emotional expression information related to depression.
    }
\item[III)]{The depression detection systems employing the ``Embedding'' representation approach surpassed those utilizing the ``0/1'' representation approach. This indicates that, compared to the straightforward ``0/1'' representation, the ``Embedding'' representation methodology encompasses more comprehensive information pertaining to the emotional expression inconsistency exhibited between the acoustic and textual modalities. As a result, it allows the model to more effectively extract crucial information for enhancing the accuracy of predicting various levels of depression severity.}
\item[IV)]{In the ``A+E'' and ``A+T+E'' configurations, the systems using supplementary ATEI information derived from the intermediate fully connected layer (FC2) achieved the optimal performance.
    It suggests that the outputs of FC2 exhibit the highest fidelity in characterizing the emotional expression inconsistency across the acoustic and textual modalities of patients with depression.}
\item[V)]{Compared to the ``A+T+E'' system using traditional emotional features derived from a single acoustic sentiment label~\cite{xuchangqing2024}, the proposed ``A+T+E'' systems with multiple sentiment labels (last three lines in TABLE~\ref{tab:integration-ATEI}) demonstrated superior subject-level accuracy. This indicates that the proposed ATEI information provides a more nuanced understanding of emotional expressions associated with depressed patients, thereby enhancing the diagnostic accuracy of depression.
    }
\end{itemize}

\subsubsection{\textbf{Integrating ATEI information}}

Furthermore, our study explored the suitable methods for integrating ATEI into the SOTA depression detection models. The performance of depression detection systems, which incorporate additional ATEI embedding features with diverse fusion strategies, is presented in TABLE~\ref{tab:ATEI-different-fusion-strategy}.
From the results presented in this table, we observed that in both the ``A+E'' and ``A+T+E'' configuration, the depression detection systems incorporating a concatenation fusion technique demonstrated superior performance compared to those utilizing addition and multiplication fusion techniques.
These systems exhibited an enhancement in subject-level accuracy, with an improvement ranging from 0.37\% to 1.1\% compared to the other two fusion methodologies.
This suggests that the concatenation fusion strategy is particularly adept at integrating the acoustic, textual, and ATEI embedding features related to depression.

\begin{table}[htbp]
\renewcommand{\arraystretch}{1.2}
\caption{The subject-level performance of the depression detection systems using additional ATEI embedding features with different fusion strategies.}
\begin{center}
    \scalebox{1}[1]{
       \begin{tabular}{c||c|cccc} \hline
       \specialrule{1.5pt}{0pt}{0pt} 
            Features & ATEI Fusion & Acc & F1 & Pre & Rec \\ \hline
             \multirow{3}{*}{A+E} &Add & 75.74 & 75.97 & 75.5 & 76.44 \\
               & Mult & 76.47 & 76.62 & 76.13 & \textbf{77.11} \\
               & Concat & \textbf{76.84} & \textbf{76.80} & \textbf{76.67} & 76.93 \\ \hline\hline
             \multirow{3}{*}{A+T+E} & Add & 78.68 & 78.46 & 78.08 & 78.85 \\
               & Mult & 79.04 & 78.87 & 78.56 & \textbf{79.19} \\
               & Concat & \textbf{79.41} & \textbf{78.90} & \textbf{79.06} & 78.74 \\ \hline
               \specialrule{1.5pt}{0pt}{0pt} 
       \end{tabular}}
\label{tab:ATEI-different-fusion-strategy}
\end{center}
\end{table}

As mentioned before, the degree of ATEI varies among individuals depending on the severity of their depression.
For example, those experiencing more profound depression would prefer to utilize a monotone voice when communicating sad textual content.
Therefore, before integrating the ATEI information within the depression modeling framework, a learnable scaling factor is introduced to adjust the degree of the ATEI embedding features. As shown in TABLE~\ref{tab:scaled-ATEI}, the incorporation of scaling factors resulted in further enhancements in system performance.
For instance, the ``A+T+E'' system utilizing depression-related acoustic, textual, and ATEI features, which employs a scaling technique during the fusion process, achieved an absolute increase of 1.84\% in accuracy compared to its counterpart that does not employ the scaling approach.
It also demonstrated a substantial subject-level accuracy improvement of 5.51\% absolute over the respective ``A+T'' baseline system.

\begin{table}[htbp]
\renewcommand{\arraystretch}{1.2}
\caption{The subject-level performance of the depression detection systems using additional ATEI embeddings with or without a scaling technique. The concatenation fusion strategy is used to integrate the ATEI embedding features.}
\begin{center}
    \scalebox{1}[1]{
       \begin{tabular}{c||c|cccc} \hline
       \specialrule{1.5pt}{0pt}{0pt} 
            Features &  Scaling & Acc & F1 & Pre & Rec \\ \hline
            A &-& 72.79 & 72.43 & 72.13 & 72.74 \\ \hline
            \multirow{2}{*}{A+E} &  $\times$ & 76.84 & 76.8 & 76.67 & 76.93 \\
                  & \checkmark & \textbf{77.94} & \textbf{77.89} & \textbf{77.48} & \textbf{78.31} \\ \hline\hline
             A+T &-& 75.74 & 75.31 & 74.95 & 75.67   \\\hline
             \multirow{2}{*}{A+T+E}   & $\times$ & 79.41 & 78.9 & 79.06 & 78.74 \\
                &  \checkmark & \textbf{81.25} & \textbf{80.97} & \textbf{80.63} & \textbf{81.31} \\ \hline
                \specialrule{1.5pt}{0pt}{0pt} 
       \end{tabular}}
\label{tab:scaled-ATEI}
\end{center}
\end{table}



\begin{figure}[ht]
    \centering
    \includegraphics[width=3.5in]{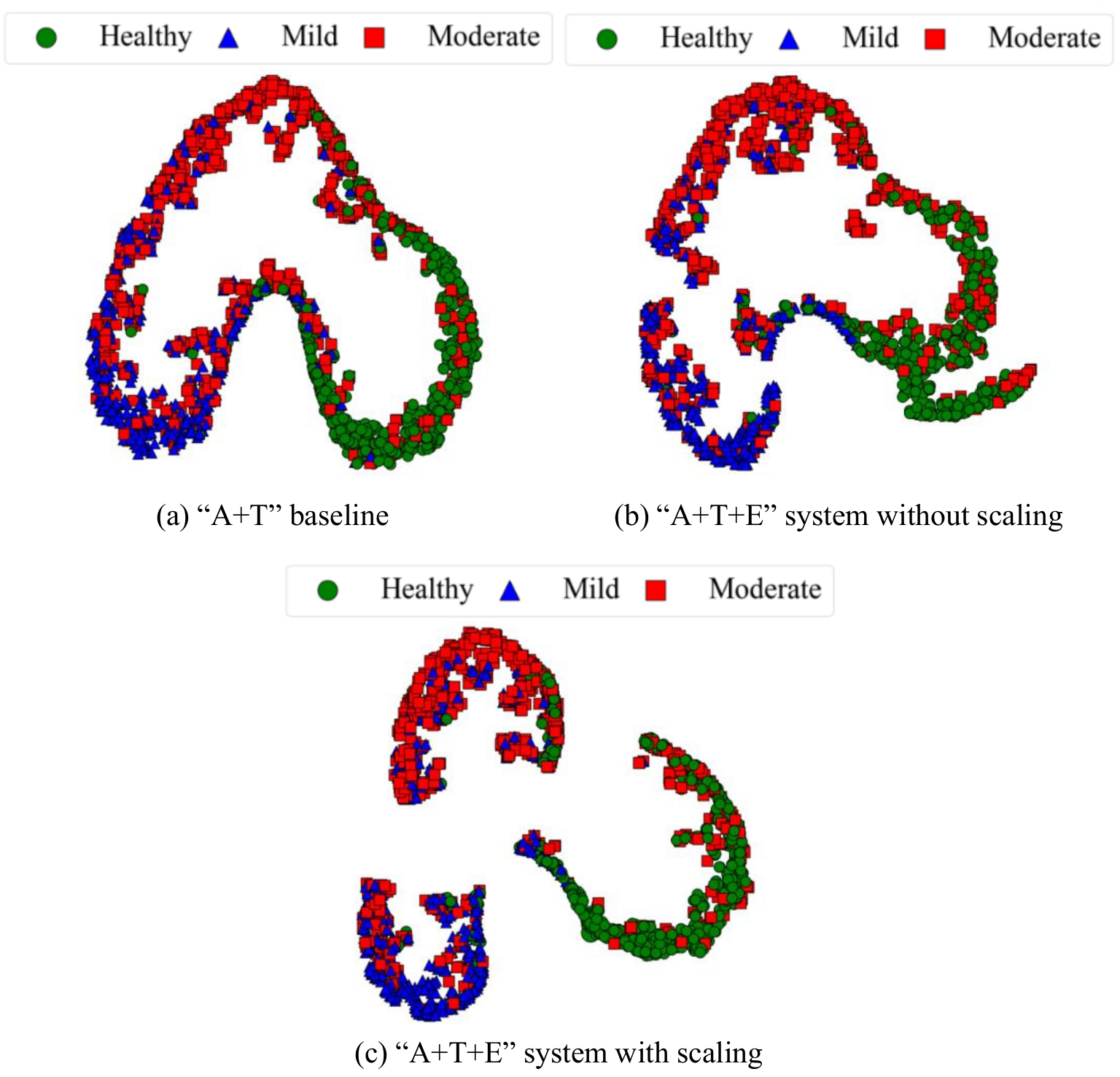}
    \caption{The t-SNE projection was performed on the outputs from the final hidden layer of the depression detection systems in TABLE~\ref{tab:scaled-ATEI}: (a) ``A+T'' depression detection baseline, (b) ``A+T+E'' system incorporating ATEI embedding features without scaling, (c) ``A+T+E'' system incorporating ATEI embedding features with scaling. The ATEI embedding features of (b) and (c) were derived from the middle fully connected layer (FC2) of Fig.~\ref{fig:ATEI-cues-extraction}. The points depicted in green, blue, and red correspond to the examples from the healthy controls, individuals with mild depression, and those with moderate depression, respectively. It is better to see it in color.}
    \label{fig:t-SNE}
\end{figure}
Finally, to more intuitively illustrate the effectiveness of the proposed depression detection framework of Fig.~\ref{fig:proposed-framework}, we employed the t-SNE technique to analyze the outputs from the final hidden layer of the depression detection systems in the second part of TABLE~\ref{tab:scaled-ATEI}.
As shown in Fig.~\ref{fig:t-SNE}, compared to the ``A+T'' baseline system, the introduction of ATEI information ((b) and (c)) allows for a more distinct separation of sample distributions corresponding to different groups, including healthy individuals and those with mild and moderate depression.
This suggests that the depression identification information is well preserved in the ATEI representations generated by the multimodal cross-attention method, which provides complementary information to the acoustic and textual features associated with depression.
Comparing (b) and (c) in Fig.~\ref{fig:t-SNE}, we found that incorporating an additional scaling technique during the fusion process can further enhance the distinct separation of sample distributions across varying degrees of depression severity. This suggests that a scaling technique can facilitate the model's ability to extract feature representations that are highly correlated with the severity of depression, thereby further improving detection accuracy.

\section{Conclusion}
\label{sec:conclusion}
According to the ECI theory in psychology, depressed patients exhibit emotional blunting. This results in inconsistent emotional expressions during natural conversations, which are abundant in the counseling conversational data. Effectively extracting this emotional expression inconsistency from counseling conversations and applying it into modeling is crucial for automatic depression detection.
This paper presented a Transformer-based framework using additional ATEI information for predicting the severity of depression.
In this framework, a multimodal cross-attention method captures the ATEI information by analyzing the complex local and long-term dependencies of the emotional expression across both the acoustic and textual domains. Additionally, various fusion techniques are used to integrate this ATEI information into SOTA depression detection methods and a scaling method is employed to adjust the degree of the ATEI embedding features during the fusion process.
Experimental results showed that the depression detection systems with additional ATEI information demonstrated significantly higher accuracy compared to their respective baseline systems lacking such information, resulting in absolute subject-level accuracy improvements of 0.73\%-5.51\%. It is interesting that the mere integration of supplementary ATEI information, encoded as ``0/1'', has the potential to enhance the efficacy of the pre-existing depression evaluation framework. Specifically, the best system was obtained by using the ATEI information represented by the ``Embedding'' vector, concatenation fusion strategy, and scaling technique. This depression detection system gave an accuracy improvement of 5.51\% absolute over the acoustic-textual baseline system. In future work, we will focus on the other techniques for extracting ATEI information.

\ifCLASSOPTIONcompsoc
  \section*{Acknowledgments}
  This work is supported by National Natural Science Foundation of China (NSFC U23B2018 and NSFC 62271477),
  Shenzhen Science and Technology Program (KQTD20200820113106007), and
  ShenZhen Fundamental Research Program (JCYJ20220818101411025, JCYJ20220818101217037 and JCYJ20220818102800001).
\else
\fi




%
\bibliographystyle{IEEEtran}
\bibliography{ref}

%
%

%








\end{document}